\documentclass[11pt]{article}
\usepackage{amsfonts,amssymb,amsmath,amsthm,bm}
\usepackage{latexsym,url,color,enumerate}
\usepackage{graphicx}

\usepackage{dirtytalk}

\newcommand\blfootnote[1]{%
  \begingroup
  \renewcommand\thefootnote{}\footnote{#1}%
  \addtocounter{footnote}{-1}%
  \endgroup
}
\usepackage{graphicx}
\usepackage[utf8]{inputenc}
\usepackage[english]{babel}
\usepackage{hyperref}
\usepackage{amsmath,bm,mathrsfs}
\usepackage{amssymb}
\usepackage{relsize}
\usepackage{mathtools}
\usepackage{bbold}
\usepackage{mathtools}
\usepackage{amsthm}
\usepackage{comment}
\usepackage{enumitem}
\usepackage[dvipsnames]{xcolor}
\usepackage{authblk}
\usepackage{IEEEtrantools}

\allowdisplaybreaks[4]

\newtheorem{theorem}{Theorem}
\newtheorem{remark}[theorem]{Remark}
\newtheorem{lemma}[theorem]{Lemma}
\newtheorem{definition}[theorem]{Definition}
\newtheorem{observation}[theorem]{Observation}

\newtheorem{example}[theorem]{Example}
\newtheorem{examples}[theorem]{Examples}
\newtheorem{corollary}[theorem]{Corollary}

 \newenvironment{indent-list}{\noindent{}\begin{itemize}}{\end{itemize}}

 \newcommand\numleq[1]%
 {\stackrel{\scriptscriptstyle(\mkern-1.5mu#1\mkern-1.5mu)}{\leq}}

 \newcommand\numeq[1]%
 {\stackrel{\scriptscriptstyle(\mkern-1.5mu#1\mkern-1.5mu)}{=}}

 \newcommand\numgeq[1]%
 {\stackrel{\scriptscriptstyle(\mkern-1.5mu#1\mkern-1.5mu)}{\geq}}

\DeclareMathOperator{\aut}{Aut}

\begin{document}

\title{Bounds on the Capacity of PIR over Graphs}
\author[$*$]{\textbf{Bar Sadeh}}
\author[$\dagger$]{\textbf{Yujie Gu}}
\author[$*$]{\textbf{Itzhak Tamo}}

\affil[$*$]{\footnotesize Department of Electrical Engineering–Systems, Tel-Aviv University, Tel-Aviv 39040, Israel.}
\affil[$\dagger$]{Faculty of Information Science and Electrical Engineering, Kyushu University, Fukuoka, Japan.}
\date{\vspace{-5ex}}
\maketitle


\begin{abstract}
	In the private information retrieval (PIR) problem, a user wants to retrieve a file from a database without revealing any information about the desired file's identity to the servers that store the database. 
	In this paper, we study the PIR capacity of a graph-based replication system, in which each file is stored on two distinct servers according to an underlying graph.
	This paper aims to provide upper and lower bounds to the PIR capacity of graphs via various graph properties. In particular,  
	we provide several upper bounds on the PIR capacity that apply to all graphs. We further improve the bounds for specific graph families (which turn out to be tight in certain cases) by utilizing the underlying graph structure.  
	For the lower bounds,  we establish optimal rate PIR retrieval schemes for star graphs via edge-coloring techniques. Lastly,  we provide an improved  PIR scheme for complete graphs, which implies an improved general lower bound on all graphs' PIR capacity.

\end{abstract}

\section{Introduction}
\blfootnote{
This work was partially supported by the European Research Council (ERC grant number 852953), NSF-BSF  (grant number 2015814), and by the Israel Science Foundation (ISF grant number 1030/15).
} 
Private information retrieval (PIR), introduced by Chor \emph{et al.} \cite{CGKS95,CKGS98}, aims at enabling users to efficiently retrieve a file from public databases while not revealing any information on the identity of the desired file (privacy requirement).  In the classical settings of~\cite{CGKS95}, there are $K$ files that are replicated across $N$ non-communicating servers and a user that generates and sends $N$ queries to the servers, one for each server. Upon receiving its query, a server responds truthfully with an answer to the user, who can then use the answers to retrieve the desired file (reliability requirement).  
 
In the PIR problem, one aims at constructing efficient retrieval schemes, where the total communication cost measures efficiency during the retrieval scheme, i.e., the total number of bits sent by the user and the servers. This was, in fact, the main figure of merit in the original paper \cite{CGKS95}, 
see also \cite{Amba1997,DG2015,WG2004}.  
On the other hand, in~\cite{CHY2015,SJ2016}, the main figure of merit was only the download cost (the cost of sending the answers by the servers), motivated by the understanding that the query size (measured in bits) is negligible when compared with the current file sizes and therefore also the answers' size. In this work, we adopt  this approach 
and define the maximum ratio between the retrieved file size and the download cost of a retrieval scheme to be the PIR capacity.

In recent years the PIR problem has attracted a tremendous amount of attention from the research community. 
Among many results, the PIR capacity in the classical settings, where each server holds a replica of the database,  was determined in \cite{SJ2016}.
Moreover, many variants of the PIR problem were also considered and studied. 
For instance, the PIR problem under colluding servers, where some servers are allowed to communicate and share the queries between themselves \cite{FGHK2017,SJ2018,SJ2016-CS};
the coded PIR problem \cite{BU2018-coded,CHY2015,TGR2018} where the files are not replicated but rather encoded using linear codes and then stored across the servers; 
the symmetric PIR problem \cite{HJ2019,WS2019} where both the user and databases' privacy should be preserved,  and many more.  

Although coding techniques for storage systems have been extensively developed in recent decades, and even some of them have found their way to real-world systems, still many system designers favor replication over coding as a means to provide resiliency to data loss due to reasons such as simplicity of implementation, file updates, and availability (see \cite{Apache2019,HDFS,RR2020}).
Nevertheless, with the current amounts of stored data,  databases are far too large to be stored on a single server. Hence, it is impractical to replicate the whole database across all servers fully \cite{RTY2020}. Therefore,  practically every server stores only a part of the entire database, and each file is replicated among several servers. 
Such a replication system is modeled by a hypergraph whose vertices represent the servers, and its hyper-edges represent the files, where a hyperedge (a file) contains the vertices (servers) that store this file. Such a system is also called a \emph{(hyper-) graph-based replication system}~\cite{RTY2020}, and if each file is stored on exactly $r$ servers, it is called an \emph{$r$-replication system}.
We note that the classical PIR settings, where each server stores the entire database, are just a special case of this problem.

In this work, we consider the PIR problem for graph-based systems,  i.e.,  a  $2$-replication system modeled by a simple graph). 
This problem and its extensions have been considered in several recent works.  
In \cite{RTY2020}, Raviv \emph{et al.} who initially introduced this model, studied the resiliency of the system against colluding servers.  
In~\cite{BU2018}, Banawan~\emph{et al.} explored the information-theoretic graph-based PIR capacity in the non-colluding model; however, their graph replication model is different from the model considered in this paper. Indeed, in \cite{BU2018} the files are the graph's vertices, and each server is an edge that connects between the files the server stores.  Hence, by assumption, each server stores \emph{only} two files, whereas no such constraint is imposed in our model. In \cite{JS2019}, Jia \emph{et al.} studied the hypergraph-based PIR problem with additional properties that both the user's privacy and the security of the stored data are protected against colluding servers.

Despite these works, the graph-based PIR problem is far from being understood, even for the simplest model of non-colluding servers. In this work, we try to bridge that gap and explore connections between this problem and the underlying graph structure. 
  In particular, we 
 derive two incomparable general upper bounds on the PIR capacity for general graphs via linear programming.
Then, we further improve these  bounds and derive tight upper bounds for star  and 
Hamiltonian vertex-transitive graphs. This is achieved by 
  utilizing the  underlying structure of the specific graph at hand. 
We complement these upper bounds with the following lower bounds. For star graphs, we construct an optimal rate PIR retrieval scheme using hypergraph edge-coloring.
Lastly, we provide an improved PIR scheme for complete graphs, which implies an improved general lower bound on the PIR capacity for all graphs.
For the reader's convenience, we summarize our results in the following table.

\begin{table}[ht]
\centering
\begin{tabular}[t]{|l|c|c|}
\hline
Graph on $N$ vertices & Capacity Bounds & Reference  \\ \hline \hline
  Star graph $S_N$  &   $\mathscr{C}(S_N) = \Theta(N^{-\frac{1}{2}})$  & Theorem~\ref{thm-lowbd-star-optimal} \\ \hline
  Complete graph $K_N$  & $ \frac{2^{N-1}}{2^{N-1}-1}\cdot \frac{1}{N} \leq \mathscr{C}(K_N) \leq \frac{2}{N+1}$ & Theorem~\ref{thm-completegraphs-imporved-lowerbd}, Corollary~\ref{coro-complete-upper} \\ \hline
  General graph $G$ & $ \mathscr{C}(K_N) \leq \mathscr{C}(G) \le \min\bigg\{\frac{\Delta}{|\mathcal{W}|}, \frac{1}{\nu(G)}, \frac{L}{X} \bigg\}  $ & Theorem~\ref{thm-uppbd-general}, Corollary~\ref{theorem-exact-neighbors-bound} \\
\hline
\end{tabular}
\caption{Summary of the main results. For the definitions and notations see Section \ref{section-problem-statment}, and for the upper bound on a general graph, see  Theorem~\ref{thm-uppbd-general} and  Corollary~\ref{theorem-exact-neighbors-bound}.} \label{table-summury}
\end{table}

The remainder of this paper is organized as follows. 
The problem statement and notations are given in Section \ref{section-problem-statment}. 
The upper bounds on the PIR capacity are provided in Section \ref{section-Bounds}. 
In Section \ref{section-star} we present a  capacity-achieving PIR scheme for star graphs. 
In Section \ref{section-complete}we construct a PIR retrieval scheme for complete graphs.
We conclude in Section \ref{sec-conclusion} with concluding remarks and some open questions.

\section{Problem statement and notations
}\label{section-problem-statment}
For positive integers $a<b$ let $[a,b]=\{a,a+1,\ldots,b\}$ and $[a]=\{1,\ldots,a\}$. For elements $A_i,i\in I$ with indices in a set $I$ and a subset $J\subseteq I$ we let $A_J=\{A_i:i\in J\}$. 

We consider the $2$-replication private information retrieval problem in the non-colluding setting, meaning that each file in the system is replicated and stored on exactly two servers, any two servers store at most one common file,  and servers do not share any information between themselves.
Let $\mathcal{S}=\{S_1,S_2,\ldots,S_N\}$ denote  $N$ non-colluding servers and $\mathcal{W}=\{W_1,$ $W_2,\ldots,W_K\}$ denote $K$ independent files. Each $W_i\in \mathbb{F}_2^L$ is a binary vector of length $L$ chosen uniformly at random from  all the vectors of $\mathbb{F}_2^L$, therefore  	
    \begin{IEEEeqnarray*}{rCl}
	    &H(W_1)=\ldots=H(W_K)=L\\
	    &H(W_1,\ldots,W_K)=H(W_1)+\cdots+H(W_K)=KL.
    \end{IEEEeqnarray*}

This model naturally can be described by a simple graph (i.e., a graph with no parallel edges and loops) $G=(\mathcal{S},\mathcal{W})$, whose vertex set is the set of servers  $\mathcal{S}$, and the edges are the files $\mathcal{W}$. A file   $W_k\in \mathcal{W}$ is also viewed as the edge $\{S_i,S_j\}$  if the file  is stored  on servers $S_i$ and $S_j$.
We denote by $W_{S_i}= \{W\in \mathcal{W}: S_i\in W\}$  the set of files stored on server $S_i$, and by $W_{i,j}$ the file that is stored on  servers $S_i,S_j$.
Hence, an instance of the PIR problem is uniquely defined by a graph, and any graph gives rise to an instance of the above  PIR problem.

A user wants to retrieve a file $W_{\theta}$ privately, where $\theta$ is chosen uniformly at random from the set  $[K]$, in the sense that each server learns nothing about the requested file index $\theta$. To that end, the user uses a random variable $\mathcal{R}$ whose realization $r$ is kept private and generates $N$ queries $Q^{[\theta,r]}_1,\ldots,Q^{[\theta,r]}_N$, one for each server. The query $Q^{[\theta,r]}_i$ is then sent  to server $S_i$. 
Throughout the paper, when it is understood from the context, we shall omit the superscript $[\theta,r]$ and simply write $Q_i$ instead of $Q^{[\theta,r]}_i$.     
Let $ \mathcal{Q}\triangleq \big\{Q_{n}: n\in [N]\big\}$ denote the set  of all queries  generated by the user.
Since the user has no information on the content of the files,  the queries are independent of them, which means that    
	\begin{equation*}
	    I(W_1,\ldots,W_K;  Q_1,\ldots,Q_N)=0. 
	\end{equation*}
Server $S_i$ upon receiving its query  $Q_i$ replies with an answer $A_i$, which is a function of the query $Q_i$ and the files it holds $W_{S_i}$. Thus, for any $i\in [N]$
	 \begin{equation*}
	     H(A_i|Q_i,W_{S_i})=0.
	 \end{equation*}
A PIR scheme has two formal requirements, \emph{reliability} and \emph{privacy}, described next. 
\vspace{0.2cm}

{\bf Reliability:}  The user should be able to  retrieve the desired file $W_{\theta}$ from the received answers $A_i$ with  zero probability of error, hence  
	 \begin{equation*}
	     H\big(W_{\theta}|A_{[N]},Q_{[N]}\big)=0.
	 \end{equation*}
     
     \vspace{0.2cm}
{\bf Privacy:} Each server learns no information about the desired file index $\theta$, i.e., for any $i\in [N]$ 
     \begin{equation*}
         H(\theta | Q_i,W_{S_i}) = H(\theta)=\log (K). 
     \end{equation*}

The main figure of merit for the PIR problem is its efficiency, i.e., how many more bits are transmitted than the actual file size being retrieved. As in~\cite{CHY2015, SJ2016}, we assume that the file size $L$ is significantly larger than the number of files in the system and therefore also than the query size, which means that     $H(W)\gg H(Q)$ for any query $Q\in \mathcal{Q}$ and file $W\in \mathcal{W}$. 
Hence, the query upload cost is negligible compared with the download cost (the answers $A_i$), and therefore we will merely focus on the download cost. 
 We define the {\it PIR rate} of a retrieval scheme $T$ as 
    \begin{equation*}
	R_{T,L} \triangleq \frac{L}{ \sum_{i\in [N]} H(A_i)},
	\end{equation*}	
which is the ratio between the retrieved file size and the total number of bits incurred in the system due to the answers. We note that we will omit the subscripts $L$ and $T$ in the sequel.   
As usual, the {\it PIR capacity} of a graph $G$ is defined as the best possible rate for arbitrarily large file size, i.e., the capacity of $G$ is defined as 
\begin{equation*}
\mathscr{C}(G)\triangleq \sup_{T,L} R_{T,L}.
\end{equation*}
\color{black}   
  
	
\section{Upper bounds for the PIR capacity} \label{section-Bounds}
In this section, we derive several upper bounds on the PIR capacity of general graphs. We begin with a bound that was originally derived in \cite{RTY2020}, however, here, we show that this bound holds more generally in the model of non-colluding servers. Then, we proceed to derive new incomparable upper bounds for general graphs. Lastly, we present an improved bound for star and  Hamiltonian vertex-transitive graphs.
	
These derived upper bounds should be compared with the known lower bounds on the PIR capacity.    
The best known lower bound for the PIR capacity of a general graph is $\mathscr{C}(G)\ge 1/N$, which follows from a retrieval scheme given in \cite[Section III]{RTY2020}. In section~\ref{section-complete} we will improve the lower bound for general graphs by improving the lower bound for complete graphs.
	
\subsection{Bounding the capacity via  linear programming}
\label{Bounding the capacity}
Before presenting the main theorem of this subsection and its proof, we recap the following standard definitions from graph theory. Let $G=(V,E)$ be a graph. 
A \textit{matching} $M\subseteq E$ of $G$ is a subset of edges, such that no two edges have a common vertex. The \textit{matching number} of $G$, denoted by $\nu(G)$ is the maximum size of a matching of $G$, i.e., $\nu(G)~=\max\{|M| : M \text{ is a matching of } G \}$.
Lastly, the degree of a vertex is the number of edges incident to it, and $\Delta$ the maximum degree of $G$  is the degree of the vertex with the greatest number of edges incident to it. 

The following theorem provides a general upper bound on the PIR capacity as a function of various graph parameters.
\begin{theorem}\label{thm-uppbd-general} 
    Let $G=(\mathcal{S},\mathcal{W})$ be a graph with maximum degree $\Delta$, then its PIR  capacity satisfies
    \begin{equation}
        \label{bound} 
        \mathscr{C}(G)\le \min\bigg\{\frac{\Delta}{|\mathcal{W}|}, \frac{1}{\nu(G)}\bigg\}.
    \end{equation}
\end{theorem} 
	
The proof of Theorem \ref{thm-uppbd-general}follows from the following observation and lemma.
\begin{observation} \label{obs-no-file-dependent}
For any answer  $A_i,i\in [N]$,  a requested  file index $k\in [K]$ and a set $J\subseteq [K]$
\begin{equation}
H(A_i|Q_i,W_J,\theta=k)= H(A_i|Q_i,W_J). \label{main-eq}
\end{equation}

\end{observation}
Indeed, given the query $Q_i$, the answer $A_i$ is a  function of the files stored on the $i$-th server, independent of the requested file $\theta=k$. 

\begin{lemma} \label{lemma-edge-bound} 
 Let $i,j \in [N]$ be two distinct servers that share the $k$-th file  $W_k\in \mathcal{W}$, then
 \begin{IEEEeqnarray*}{rCl}
        H(A_i) + H(A_j) &\geq&  H\big( A_{i}| \mathcal{W}\setminus \{W_{k}\}, \mathcal{Q} \big) 
        +\> H\big( A_{j}| \mathcal{W}\setminus \{W_{k}\}, \mathcal{Q} \big) 
        \ge L. 
	\end{IEEEeqnarray*}
\end{lemma}

\begin{IEEEproof} 
The proof goes along the same lines as  the proof of Lemma 1 in   \cite{BU2018}. The first inequality holds since conditioning reduces entropy, hence  we proceed to prove the second inequality.   
\begin{IEEEeqnarray}{rCl}
	L&=&  H(W_k)	\nonumber \\ 
    &=& H(W_k | \mathcal{W} \setminus \{W_k\} , \mathcal{Q}, \theta=k)  \label{leb1} \\
    &=& H(W_k | \mathcal{W} \setminus \{W_k\} , \mathcal{Q}, \theta=k) 
    -\> H(W_k | \mathcal{W} \setminus \{W_k\} , \mathcal{Q}, \theta=k, A_{[N]})  \label{leb2} \\
    &=& H(W_k | \mathcal{W} \setminus \{W_k\} , \mathcal{Q}, \theta=k) 
    -\> H(W_k | \mathcal{W} \setminus \{W_k\} , \mathcal{Q}, \theta=k, A_i,A_j) \label{leb3} \\
    &=& I(W_k; A_i, A_j | \mathcal{W} \setminus \{W_k\} , \mathcal{Q}, \theta=k) \nonumber \\
    &=& H( A_i, A_j | \mathcal{W} \setminus \{W_k\} , \mathcal{Q}, \theta=k)  \label{leb4} \\
    &\leq&  H(A_i  | \mathcal{W} \setminus \{W_k\} , \mathcal{Q}, \theta=k) 
    +\> H(A_j  | \mathcal{W} \setminus \{W_k\} , \mathcal{Q}, \theta=k) \nonumber \\
    &=& H(A_i  | \mathcal{W} \setminus \{W_k\} , \mathcal{Q}) + H(A_j  | \mathcal{W} \setminus \{W_k\} , \mathcal{Q}), \label{leb5}
\end{IEEEeqnarray}
where \eqref{leb1} follows since $W_k$ is independent of the remaining files, the queries and the requested file $\theta=k$; \eqref{leb2} follows from the reliability requirement, i.e., $W_k$ can be retrieved from the queries $\mathcal{Q}$, $\theta=k$ and the answers  $A_{[N]}$; \eqref{leb3} holds since  the answers  $A_{[N]} \setminus \{A_i,A_j \}$ are a function of $\mathcal{Q}$ and $\mathcal{W}\setminus \{W_k\}$; \eqref{leb4} holds since $H(A_i,A_j | \mathcal{W},\mathcal{Q}, \theta=k)=0$ and lastly, \eqref{leb5} follows from \eqref{main-eq}.
\end{IEEEproof}

Next, we provide the proof of Theorem~\ref{thm-uppbd-general}, which is similar to the proof of~\cite[Lemma 8]{RTY2020}. We note that Theorem~\ref{thm-uppbd-general} is, in fact, a strengthening of \cite[Lemma 8]{RTY2020} since we show that the same result holds without the assumptions \cite{RTY2020} that the servers may collude and share their queries amongst themselves. 

We recall the following definition from graph theory. The \textit{incidence matrix $I(G)$} of a graph $G$ is a $|V(G)| \times |E(G)|$ binary matrix in which rows correspond to vertices and columns
correspond to edges, and an entry contains $1$ if and only if the respective vertex is incident with the respective edge, otherwise it is zero. 
    \begin{IEEEproof}[Proof of Theorem~\ref{thm-uppbd-general}]
    Consider the rate of some PIR scheme for $G$  
    \begin{IEEEeqnarray*}{rCl}
        R &=& \frac{L}{\sum_{i\in [N]} H(A_i)}=
        \frac{1}{\sum_{i\in [N]} (H(A_i)/L)} 
        =
        \frac{1}{\sum_{i\in [N]} \mu_i}
        =\frac{1}{\mathbb{1}_N\cdot \boldsymbol{\mu}^T},
    \end{IEEEeqnarray*}
    where $\mu_i\triangleq H(A_i)/L$ for any $i\in [N]$, $\boldsymbol{\mu}\triangleq (\mu_1,\cdots,\mu_N)$, and $\mathbb{1}_N$ is the all-ones vector of length $N$. 
    According to Lemma~\ref{lemma-edge-bound}, if servers $i$ and $j$ are adjacent, then $\mu_i+\mu_j\ge 1$. Hence, we can get an upper bound on the PIR rate via the reciprocal of the optimal value of the following linear program. 
    \begin{equation} \label{orig-linear-problem}
		\min \; \ {\mathbb{1}_N\cdot \boldsymbol{\mu}^T}, \; 
		\text{subject to }
		I(G)^T\cdot \boldsymbol{\mu}^T \geq \mathbb{1}_K \; \text{and}\; \boldsymbol{\mu}\geq 0.
	\end{equation}
		Its dual problem is 
	\begin{equation} \label{dual-linear-problem}
		\max \; \ {\mathbb{1}_K\cdot \boldsymbol{\eta}^T}, \; 
		\text{subject to }
		I(G)\cdot \boldsymbol{\eta}^T \leq \mathbb{1}_N \; \text{and}\; \boldsymbol{\eta}\geq 0,
	\end{equation}	
	where $\boldsymbol{\eta}$ is a vector  of length  $K$.
	By  the primal-dual theory, any feasible solution to~\eqref{dual-linear-problem} is a lower bound for~\eqref{orig-linear-problem}.
	Below we provide two feasible solutions to~\eqref{dual-linear-problem}.
	\begin{itemize}
	    \item[(S1)]  It is readily verified that  $\boldsymbol{\eta}=\frac{1}{\Delta} \cdot \mathbb{1}_K$ is a  feasible solution of~\eqref{dual-linear-problem}, where $\Delta$ is the maximum degree in $G$. Therefore, $\ {\mathbb{1}_K\cdot \boldsymbol{\eta}^T}=K/\Delta$, which implies that the PIR rate is at most  $\Delta/K$,  
	    \item[(S2)] 
	    Let $M\subseteq \mathcal{W}$ be a maximum matching of $G$, i.e.,  $|M|=\nu(G)$. Let   $\boldsymbol{\eta}=(\eta_1,\dots,\eta_K)\in \{0,1\}^K$ be an indicator vector of the set $M$, i.e.,  $\eta_i=1$ if and only if $W_i\in M$. Again, it is readily seen that $\boldsymbol{\eta}$ is a feasible solution of~\eqref{dual-linear-problem} and therefore as before the PIR rate is at most  $1/\nu(G)$. 
	\end{itemize}
	This completes the proof of the two upper bounds.
\end{IEEEproof}
We now turn to present different types of graphs and their capacity bounds obtained by Theorem~\ref{thm-uppbd-general}. The examples show that the two upper bounds from Theorem~\ref{thm-uppbd-general} are incomparable in general, as seen in the first and last examples. 
\begin{examples}
\begin{enumerate}
    \item[] 
    \item[1)] Let $G$ be a regular graph of degree $\Delta$ with an odd number of vertices $N$, and $K=\frac{\Delta \cdot N}{2}$ edges. It follows that $\nu(G) \leq \frac{N-1}{2}$. Hence
      \begin{equation*}
        \frac{1}{N}\leq \mathscr{C}(G)\le min \{\frac{2}{N}, \frac{1}{\nu(G)} \} = \frac{2}{N}< \frac{1}{\nu(G)}.  
      \end{equation*}
      Note that this upper bound is, in fact, tight for some regular graphs, as can be verified by a graph that is a perfect matching on $N$ vertices for an even $N$. However,   this upper bound is not tight in general for regular graphs, and in Section \ref{Hamiltonian-vertex-transitive}we provide a tighter bound for a specific type of regular graphs.
    \item[2)] Let $G=K_{N,M}$ be the complete bipartite graph over two sets of vertices of size  $N$ and $M$ ($N\leq M$). 
    Then, $K=N\cdot M$, $\Delta=M, \nu(G) = {N}$ and 
     \begin{equation*}
     \frac{1}{M+N}\leq \mathscr{C}(G)\le   \frac{1}{N}.
     \end{equation*}
     It is an interesting open question to determine the exact PIR capacity of this graph. 
    \item[3)] Let $G=W_{2N}$ be the wheel graph on $2N$ vertices, formed by connecting a single vertex to all vertices of a cycle of length $2N-1$. It is easy to verify that  $K=4N-2$, $\Delta = 2N-1$ and $\nu(G) = N$. Hence  
    \begin{equation*}
    \frac{1}{2N}\leq \mathscr{C}(G)\le min\{ \frac{1}{2},\frac{1}{N}\} = \frac{1}{N}.
    \end{equation*}
   This lower bound is not tight, since by merging a capacity achievable scheme for the cycle graph with the scheme for the star graph given in Theorem \ref{thm-lowbd-star-optimal} we get a PIR scheme with a rate better than $\frac{1}{2N}$.
\end{enumerate}
\end{examples}

\subsection{Bounding the PIR capacity via a system of inequalities}
The bounds given in Section \ref{Bounding the capacity}on the PIR capacity are by no means optimal and tight in the general case. Therefore, one needs to use further properties of the specific graph at hand to obtain tighter bounds.
In this section, we lower bound the amount of information sent by a specific server by some function of the amount of information sent by its neighbors. This allows us to describe an optimization problem that upper bounds the scheme rate and whose solution depends on the specific graph structure.

Then, we will focus on a specific graph, the star graph, and apply the optimization problem to improve the upper bound for the star graph obtained using Theorem \ref{thm-uppbd-general}. 
The reason for considering the star graph is twofold. First, the star graph has a very simple structure, and it is one of the simplest types of graphs for which determining the PIR capacity is still non-trivial. Second, the results developed in this section may be used for general graphs since any graph is a union of star graphs of possibly different sizes. Indeed, every vertex and its set of neighbors form a star graph.    

The main result of this section follows from the following theorem. 
\begin{theorem} \label{theorem-vetex-neighbors}
		Given a PIR scheme for a graph $G=(\mathcal{S},\mathcal{W})$, then for every server $S\in \mathcal{S}$ of degree $\delta$ and  neighbors set   $N(S)=\{S_1, \ldots, S_\delta \}$ the following holds 
		 \begin{equation*}
		     H(A_S)\geq \sum_{i=1}^{\delta} \max \bigg\{0,\,  L-\sum_{j=i}^{\delta} H(A_{S_j}) \bigg\}.
		 \end{equation*}
        It immediately follows that the best  lower bound is attained when the  $H(A_{S_j})$ are  ordered in a  descending order, i.e., $H(A_{S_1})\geq H(A_{S_2})\geq \ldots \geq H(A_{S_\delta})$. 
\end{theorem}

\begin{IEEEproof}
For  $i\in[\delta]$ let $W_i$ be the file stored on  $S$ and $S_i$. The answer $A_S$ from server $S$ satisfies 
    \begin{align}
         H(A_S) &\geq H(A_S | \mathcal{Q})   \nonumber  \\
         & = H(A_S | \mathcal{Q}, \mathcal{W} \setminus W_{[\delta]}) \label{file-condition} \\ 
              &\ge 
        I(A_S;W_{[\delta]}| \mathcal{Q}, \mathcal{W} \setminus W_{[\delta]})         		\nonumber  \\ 
        \label{eq-I-chain}   &=
        \sum_{i=1}^{\delta} I(A_S; W_i|\mathcal{Q}, \mathcal{W} \setminus W_{[\delta]}, W_{[i-1]})\\
        &= \sum_{i=1}^{\delta} \Big( H(A_S|\mathcal{Q}, \mathcal{W} \setminus W_{[\delta]}, W_{[i-1]}) - H(A_S|\mathcal{Q}, \mathcal{W} \setminus W_{[\delta]}, W_{[i]}) \Big) \nonumber \\	
    & = 	  \sum_{i=1}^{\delta} \Big( H(A_S|\mathcal{Q}, \mathcal{W} \setminus W_{[\delta]},W_{[i-1]},\theta=i) - H(A_S|\mathcal{Q}, \mathcal{W} \setminus W_{[\delta]},W_{[i]},\theta=i) \Big) \label{cond}	\\
    & =  	 \sum_{i=1}^{\delta} I(A_S;W_i|\mathcal{Q}, \mathcal{W} \setminus W_{[\delta]}, W_{[i-1]}, \theta=i )\nonumber\\
        &=   \sum_{i=1}^{\delta} \Big( H(W_i|\mathcal{Q}, \mathcal{W} \setminus W_{[\delta]},W_{[i-1]}, \theta=i) - H(W_i | A_S,\mathcal{Q}, \mathcal{W} \setminus W_{[\delta]},W_{[i-1]}, \theta=i) \Big) \nonumber
        \\
        \label{eq-to-bound-I}
        &=          \sum_{i=1}^{\delta} L -H(W_i|A_S,\mathcal{Q}, \mathcal{W} \setminus W_{[\delta]},W_{[i-1]}, \theta=i) 					\\
        \label{eq-to-bound-II}
        &=          \sum_{i=1}^{\delta} \max \bigg\{0,L -H(W_i|A_S,\mathcal{Q}, \mathcal{W} \setminus W_{[\delta]},W_{[i-1]}, \theta=i) \bigg\} 		
    \end{align}
where~\eqref{file-condition} follows since $A_S$ is independent of $\mathcal{W} \setminus W_{[\delta]}$;~\eqref{eq-I-chain} follows from the chain rule for mutual information; \eqref{cond} follows from \eqref{main-eq};
\eqref{eq-to-bound-I} follows since  $W_i$ is independent of $\{W_{[i-1]}, \mathcal{Q}, \mathcal{W} \setminus W_{[\delta]}, \theta \}$ and \eqref{eq-to-bound-II} follows since the entropy of $W_i$ given other random variables is at most $L$.  
 
Next, we turn to derive an upper bound on $H(W_i|A_S, \mathcal{Q}, \mathcal{W} \setminus W_{[\delta]}, W_{[i-1]},\theta)$. For every $i\in[\delta]$ let $A_{i:\delta}=\{ A_{S_i} , \ldots A_{S_\delta} \}$.
    \begin{IEEEeqnarray}{rCl}
    \IEEEeqnarraymulticol{3}{l}{
    H(W_i|A_S,\mathcal{Q}, \mathcal{W} \setminus W_{[\delta]},W_{[i-1]},\theta=i)
            } \nonumber\\*
        &\le & 
        H(W_i|A_S ,\mathcal{Q}, \mathcal{W} \setminus W_{[\delta]},W_{[i-1]},\theta=i, A_{i:\delta}) 
         \nonumber\\* &&+\>
        H(A_{i:\delta}|A_S,\mathcal{Q}, \mathcal{W} \setminus W_{[\delta]},W_{[i-1]},\theta=i) \label{eq-H-XYZ}\\
        &=& 
        H(A_{i:\delta} | A_S,\mathcal{Q}, \mathcal{W} \setminus W_{[\delta]},W_{[i-1]},\theta=i)\label{eq-H=0}\\
        &\le & 
        \sum_{j=i}^{\delta} H(A_{S_j}),	\label{eq-to-bound-H}
    \end{IEEEeqnarray}
    where~\eqref{eq-H-XYZ} follows from the inequality $H(X|Y)\le H(X|Y,Z) + H(Z|Y)$; 
    \eqref{eq-H=0} follows from the reliability requirement of the PIR scheme  and \eqref{eq-to-bound-H} holds since  conditioning reduces entropy and by the chain rule. Finally, by combining~\eqref{eq-to-bound-II} and \eqref{eq-to-bound-H} the result follows. 
\end{IEEEproof}

The following corollary follows immediately from Theorem \ref{theorem-vetex-neighbors}, and it provides a general bound on the PIR capacity of graphs. 
For a server $S\in \mathcal{S}$ of degree $\delta_S$ let $u^S_1, \ldots, u^S_{\delta_S}$ be its neighbors.
\begin{corollary}\label{theorem-exact-neighbors-bound}
    The PIR capacity of a graph $G=(\mathcal{S},\mathcal{W})$ is at most  $\mathscr{C}(G) \leq L/X$, where $X$ is the solution of the following minimization problem
    \begin{IEEEeqnarray}{C}
		\min \; \ {\sum_{S\in \mathcal{S}} x_S}, \;  
		\text{subject to} 
		\quad  
		x_S \geq \sum_{k=1}^{\delta_S} \max 
		\bigg\{0,\,  L-\sum_{j=k}^{\delta_S} x_{u^S_j} \bigg\} 
		\;
		\text{for every server $S\in \mathcal{S}$}.
		\label{neighbors-linear-problem} 
    \end{IEEEeqnarray}
\end{corollary}

\begin{IEEEproof}
Consider  a PIR scheme for $G$, then by   Theorem~\ref{theorem-vetex-neighbors} one can verify that the $ H(A_S)$'s for $S\in \mathcal{ S}$ are a feasible solution for \eqref{neighbors-linear-problem} and therefore $X\leq \sum_{S\in \mathcal{S}} H(A_S)$. Then 
\begin{equation*}
    \mathscr{C}(G) = \frac{L}{\sum_{S\in \mathcal{S}} H(A_S)} \leq \frac{L}{X},    
\end{equation*}
and the result follows. 
\end{IEEEproof}

Solving the minimization problem \eqref{neighbors-linear-problem} might be difficult for general graphs. Therefore, obtaining an informative (closed-form) upper bound for the PIR capacity might not be possible. Hence, to obtain a closed-form upper bound, it might be easier to apply the result of    Theorem~\ref{theorem-vetex-neighbors} for specific graph families, as we shall do next for the star graph.

A {\it star } (graph) $S_N=(\mathcal{S},\mathcal{W})$ is a simple graph on $N$ vertices with one vertex, say $S_N$, with degree $\deg(S_N)=N-1$ and the other $N-1$ vertices are of degree $1$. Hence,  $K=|\mathcal{W}|=N-1$ and  $W_i=\{S_i, S_N\}$ for $i\in [K]$. The 
 matching number is $\nu({S_N})=1$ and the maximal degree $\Delta=N-1$, then by  Theorem~\ref{thm-uppbd-general} $\mathscr{C}(S_N)\le 1$, which is uninformative, since the PIR capacity of any graph is at most $1$. 
In the next theorem, we invoke  Theorem~\ref{theorem-vetex-neighbors} to obtain a significantly tighter upper bound on the star graph's PIR capacity. In fact, the bound will be shown to be tight up to a multiplicative factor via an achievable scheme (Theorem~\ref{thm-lowbd-star-optimal}).

\begin{theorem} \label{thm-uppbd-star}
	The PIR capacity of the star graph $S_N$ satisfies $\mathscr{C}(S_N) \leq O(N^{-\frac{1}{2}})$.
\end{theorem}
\begin{IEEEproof}
    Consider a star based PIR retrieval scheme with $N$ servers and $K=N-1$ files.  Let $S_N$ be the unique vertex of degree $N-1$, and  without loss of generality  assume that  $H(A_1)\ge H(A_2)\ge \ldots \ge H(A_{N-1})$. Let $t$ be the largest integer in $[N-1]$ such that  $H(A_{N-t})\le \frac{L}{t}$, and note that $t\geq 1$, since $H(A_{N-1})\leq L$. Then, by  Theorem~\ref{theorem-vetex-neighbors} 
    \begin{IEEEeqnarray}{rCl}
        H(A_N)&\geq& \sum_{i=1}^{N-1} \max \Big\{0,\,  L-\sum_{j=i}^{N-1} H(A_j) \Big\} \nonumber\\
        &\geq& \sum_{i=N-t}^{N-1} \max \Big\{0,\,  L-\sum_{j=i}^{N-1} H(A_j) \Big\} \nonumber\\
        &=& 
        tL-\sum_{i=N-t}^{N-1}\sum_{j=i}^{N-1} H(A_{N-t})\nonumber\\
        &\geq& tL-\frac{t(t+1)}{2}\cdot\frac{L}{t}\label{second-inequality}\\
        &=& 
        tL-\frac{t+1}{2}L,\label{fifth-inequality}
    \end{IEEEeqnarray}
    where  \eqref{second-inequality} follows by the definition of $t$.
    Notice that also by the definition of $t$ that    
    \begin{equation}
        H(A_1)\geq H(A_2)\geq \ldots \geq H(A_{N-(t+1)})>\frac{L}{t+1}.\label{third-inequality}
    \end{equation} 
    Then, by combining \eqref{fifth-inequality} and \eqref{third-inequality} 
    \begin{IEEEeqnarray}{rCl}
        \sum_{i=1}^N H(A_i) 
        &\ge &
        H(A_N) + \sum_{i=1}^{N-t-1} H(A_i) \nonumber\\
        &\ge & tL-\frac{t+1}{2}L + \frac{N-t-1}{t+1}L\nonumber\\
        &\ge & L\cdot \min_{t\in [N-1]}\bigg\{ \frac{t}{2} +\frac{N}{t+1}-\frac{3}{2}\bigg\}\nonumber\\
        &\ge & \big(\sqrt{2N}-2\big)L\nonumber,
    \end{IEEEeqnarray}
    where the last inequality follows since the function  $f(x)=x/2+N/(x+1)-3/2$ attains its minimum at  $x_{\min}=\sqrt{2N}-1$. Therefore the PIR rate satisfies 
    \begin{equation*}
        \frac{L}{\sum_{i=1}^N H(A_i)}
        \le 
        \frac{L}{\big(\sqrt{2N}-2\big)L}
        =
        \frac{1}{\sqrt{2N}-2},
    \end{equation*}
    as required.   
\end{IEEEproof}

\subsection{Bounding the capacity for Hamiltonian vertex-transitive graphs} \label{Hamiltonian-vertex-transitive}
This section proves an improved upper bound on the PIR capacity for Hamiltonian vertex-transitive graphs, which turns out to be tight for  cycle graphs. We first recall the needed definitions. A graph $G$ is called {\it Hamiltonian} if it possesses a cycle that traverses every vertex in the graph, and it is called  {\it vertex-transitive graph} if its automorphism group $\aut(G)$ is transitive on its vertices, i.e., for any two vertices $u,v\in V(G)$ there exists an automorphism $f\in \aut(G)$ such that $f(v)=u$.

By Theorem~\ref{thm-uppbd-general}, the PIR capacity of the family of regular graphs on $N$ vertices  (which also includes the family  vertex-transitive graphs on $N$ vertices) is at most $ 2/N$. 
In the following theorem, we prove a slightly tighter bound. 

\begin{theorem}\label{thm-upper-vertex-trans}
Let $G=(\mathcal{S},\mathcal{W})$ be a Hamiltonian vertex-transitive graph, then 
\begin{equation}
    \mathscr{C}(G)\le \frac{2}{N+1}.  \label{Hamiltonian}
\end{equation}
\end{theorem}

Before proceeding to prove Theorem~\ref{thm-upper-vertex-trans}, we remark the following. First,  the upper bound \eqref{Hamiltonian} is tight  for certain graphs. Indeed, it was shown in \cite{BU2018} that the PIR capacity of the cycle graph $C_N$ on $N$ vertices is $\mathscr{C}(C_N)= \frac{2}{N+1}$. 
Second, since the complete graph is vertex-transitive, we have the following immediate corollary, which provides a slightly tighter 
upper bound than the one obtained by  Theorem~\ref{thm-uppbd-general} for  complete graphs. In Section~\ref{section-complete} we provide  PIR schemes (lower bounds) for complete graphs.

\begin{corollary}\label{coro-complete-upper}
\label{remark-complete-graph}The capacity of the complete graph $K_N$ on $N$ vertices satisfies 
\begin{equation*}
\mathscr{C}(K_N)\leq \frac{2}{N+1}.
\end{equation*}

\end{corollary}
To prove Theorem \ref{thm-upper-vertex-trans} we will need  the following lemma. 

\begin{lemma} \label{lemma-symetric-scheme}
In a vertex-transitive graph-based replication system, for any achievable rate $R$, there exists a scheme with a rate $R$ that satisfies
    \begin{align}
       H\big(A_{i}| \mathcal{Q}\big) = H\big(A_{j}| \mathcal{Q}\big) , & \ \ \forall\, i,j\in [N], \label{lss-1} \\ 
        H\big(A_i|\mathcal{W}\setminus \{W_{i,j}\}, \mathcal{Q}\big) \geq \frac{L}{2} ,& \ \ \forall\, i,j\in [N].\label{lss-2}
    \end{align}
\end{lemma}
The proof of Lemma~\ref{lemma-symetric-scheme} is deferred to Appendix \ref{proof-lemma-symetric-scheme}. 

\vspace{0.2 cm}
\begin{IEEEproof}[Proof of Theorem~\ref{thm-upper-vertex-trans}] 
Let $(S_1,W_1,S_2,W_2,\ldots,S_{N},$ $W_{N},S_{1})$ be a Hamiltonian cycle in $G=(\mathcal{S},\mathcal{W})$. 
By Lemma~\ref{lemma-symetric-scheme}, if $R$ is an  achievable rate, then there exists a PIR scheme with $R$ satisfying also \eqref{lss-1} and \eqref{lss-2}. Then
\begin{IEEEeqnarray}{rCl}
    L    &=& H(W_{N})                \nonumber    \\ 
    &=& H(W_{N}| \mathcal{Q}, \theta=N)        \label{eq-pf-uvt-1-1} \\ 
    &=& H(W_{N}| \mathcal{Q}, \theta=N) - H(W_{N}| A_{[N]}, \mathcal{Q}, \theta=N) \label{eq-pf-uvt-1-2} \\    
    &=& I\big(W_{N};A_{[N]}| \mathcal{Q}, \theta=N\big)   \nonumber \\    
    &=& H(A_{[N]}| \mathcal{Q},\theta=N) - H(A_{[N]}| W_{N},\mathcal{Q},\theta=N) \nonumber \\  
    &\leq& \sum_{i\in[N]} H(A_i| \mathcal{Q},\theta=N) - H(A_{[N]}| W_{N},\mathcal{Q},\theta=N) \nonumber \\ 
    &=& \sum_{i\in[N]} H(A_i| \mathcal{Q}) - H(A_{[N]}| W_{N},\mathcal{Q},\theta=N) \label{eq-pf-uvt-1-3} \\ 
    &=& N H(A_1| \mathcal{Q}) -  H(A_{[N]}| W_{N},\mathcal{Q},\theta=N), \label{eq-pf-uvt-1-4}
\end{IEEEeqnarray}
    where~\eqref{eq-pf-uvt-1-1} follows since  the queries $\mathcal{Q}$ and the identity of the required file are independent of the required file content; \eqref{eq-pf-uvt-1-2} follows from the reliability requirement of the PIR scheme; \eqref{eq-pf-uvt-1-3} follows from Observation~\ref{obs-no-file-dependent}; and~\eqref{eq-pf-uvt-1-4} is due to~\eqref{lss-1} in Lemma~\ref{lemma-symetric-scheme}.

   Rearranging the above inequality yields 
\begin{IEEEeqnarray}{rCl}
    N H(A_1| \mathcal{Q}) &\geq& L + H(A_{[N]} | W_{N},\mathcal{Q},\theta=N)   \nonumber \\* 
     & \ge& L +     H\big(A_{[N-1]}| W_{N},\mathcal{Q} ,\theta=N\big)
     \label{eq-pf-uvt-2-1}    \\
    & = & L +
    \sum_{j=1}^{N-1} 
    H\big(A_{j}| W_{N},\mathcal{Q}, A_{[j-1]}, \theta=N\big)
    \label{eq-pf-uvt-2-2}     \\
    &\ge & L +
    \sum_{j=1}^{N-1} 
     H\big(A_{j}| W_{N},\mathcal{Q}, A_{[j-1]}, \mathcal{W}\setminus\{W_{j}\} ,\theta=N\big)
     \label{eq-pf-uvt-2-3}     \\
    &\ge &L +     \sum_{j=1}^{N-1} 
    H\big(A_{j}| W_{N},\mathcal{Q},  \mathcal{W}\setminus\{W_{j}\} ,\theta=N \big)
    \label{eq-pf-uvt-2-4} \\
    &= &L + \sum_{j=1}^{N-1}  
     H\big(A_{j}| \mathcal{Q},  \mathcal{W}\setminus\{W_{j}\},\theta=N\big) \label{eq-pf-uvt-2-5}     \\
    &\ge& L + \sum_{j=1}^{N-1} \frac{L}{2} \label{eq-pf-uvt-2-6} \\*
      &=&
     \frac{(N+1) L}{2}, \nonumber
\end{IEEEeqnarray}
where~\eqref{eq-pf-uvt-2-1} follows from the non-negativity of entropy;~\eqref{eq-pf-uvt-2-2} follows from the chain rule of entropy;~\eqref{eq-pf-uvt-2-3} follows since conditioning reduces entropy;~\eqref{eq-pf-uvt-2-4} follows since  
the file  $W_{j}$ is stored only on servers $S_j,S_{j+1}$ and therefore the answers  $A_{[j-1]}$ are deterministic functions of $\mathcal{
Q}$ and $\mathcal{W}\setminus \{W_{j}\}$;~\eqref{eq-pf-uvt-2-5} follows from the fact that $W_{N}\in \mathcal{W}\setminus\{W_{j}\}$ for $j\in [N -1]$; and~\eqref{eq-pf-uvt-2-6} follows from~\eqref{lss-2} in Lemma~\ref{lemma-symetric-scheme} and Observation~\ref{obs-no-file-dependent}.
Therefore,  the PIR rate satisfies 
\begin{equation*}
  R=  \frac{L}{\sum_{i\in [N]}H(A_i)} \le
   \frac{L}{\sum_{i\in [N]}H(A_i | \mathcal{Q})} =  
   \frac{L}{NH(A_1 | \mathcal{Q})}
   \le \frac{2}{N+1}.  
\end{equation*}
\end{IEEEproof}

\begin{remark}
Although the result seems to apply to very limited graph families,  i.e., the graph's requirements are very strong, this is not precise. The question of whether every connected vertex-transitive graph has a Hamiltonian path or cycle is a famous open problem in graph theory, which is sometimes referred to as Lov\'{a}sz conjecture (see \cite{lovasz}). In fact, all the known examples of connected vertex-transitive graphs are Hamiltonian, except for four known examples: the Petersen graph, the Coxeter graph, and two graphs which are some variants of these two. 
\end{remark}


\section{Retrieval schemes for star-based PIR system} \label{section-star}
	In this section, we present retrieval schemes for the star graph. We begin with a simple scheme (Theorem \ref{thm-lowbd-star-simple})  with rate $\Omega(N^{-1})$.  Then, building on the ideas given in the first scheme, we proceed to present a scheme with an improved rate, which is, in fact, optimal up to a multiplicative factor. More precisely, the scheme has a rate  $\Omega(N^{-1/2})$ which matches the upper bound given in Theorem~\ref{thm-uppbd-star}. 
	The precise rate achieved by the first scheme is stated in the next theorem.
	\begin{theorem}\label{thm-lowbd-star-simple}
	The star graph's PIR capacity $S_N$ is at least $\mathscr{C}(S_N)\ge 2/N$.
	\end{theorem}
	
	To construct the retrieval scheme, we shall use the notion of an {\it edge-coloring} of a graph. 
	A (proper)  edge-coloring of a graph with possibly self-loops is a coloring of the graph edges, such that incident edges (i.e., edges that share a common vertex)  have distinct colors. Formally, given a graph $G=(V,E)$, an edge-coloring of it with $T$ colors is a mapping  $c: E\to [T]$ such that for any vertex $i\in V$ and $j,j'$ two distinct neighbors  of   $i$, $c(i,j)\neq c(i,j')$. Clearly, any edge-coloring requires a number of colors which is at least the maximum degree of $G$. 
	
 We will make use of the well-known Baranyai's Theorem given below. 
	
\begin{theorem}[Baranyai's theorem]\cite{Baran75,SB79} 
	\label{baranyai-theorem} 
	Let $X$  be a set of size $N$. Given $t|N$, there exist $\binom{N-1}{t-1}$ partitions of $X$ into $t$-sets such that each $t$-subset of $X$ occurs in exactly one of these partitions. 
	\end{theorem}
    \begin{remark}
    For the special case of  $t=2$ and even $N$, Theorem \ref{baranyai-theorem}provides a partition of the edges of the complete graph on $N$ vertices to $N-1$ perfect matchings.
    
    \end{remark}
 	Let $G$ be the complete graph on $N$ vertices (including self-loops), i.e., each vertex is connected to all other vertices, including itself; therefore, the degree of each vertex is $N.$
	We will show next by invoking Theorem \ref{baranyai-theorem}that $G$ admits an edge-coloring with exactly $N$ colors, which is clearly optimal.
 	\begin{lemma} \label{lemma-complete-coloring}
		The graph  $G$ admits an edge-coloring with $N$ colors.
	\end{lemma}
\begin{IEEEproof} 
	Assume first that the number of vertices $N$ is even. By Theorem \ref{baranyai-theorem} the non-loop edges of  $G$ can be partitioned into $N-1$ perfect matchings, $M_1,\ldots,M_{N-1}$. These partitions define the color classes of the edge-coloring as follows. Color the edge $(i,j)$ for $i\neq j$ by the color $l$ if matching $M_l$ is the unique matching that contains this edge. Next, color all the self-loops by the color $N$. It is easy to verify that, indeed, this is a proper edge-coloring of $G$.
	
	Next, assume that $N$ is odd, and consider the complete graph (without self-loops) with vertex set  $[N+1]$. Again, by Theorem \ref{baranyai-theorem} the edges of the graph can be partitioned to $N$ perfect matchings $M_i$. Define the coloring of $G$  as follows. For distinct vertices $i,j$ of $G$ color the edge $(i,j)$ by the color $l$ if matching $M_l$ is the unique matching that contains this edge. Lastly, for $i=1,\ldots,N$ color the self-loop  $(i,i)$  by color $l$ if edge $(i,N+1)$ is in matching $M_l$.   
	It is easily verified that this is indeed a proper edge-coloring of $G$, which completes the proof. 
	\end{IEEEproof}
	
	\begin{IEEEproof}[Proof of Theorem~\ref{thm-lowbd-star-simple}]
	The result will follow from a PIR scheme with the claimed rate, given next. 
	Suppose that each file $W=(W(1),\cdots,W(K))\in \mathbb{F}^K_2$ is a binary vector of length $K=N-1$. 
	Let $G=(V,E)$ be the complete graph (including self-loops) on vertex set $V=[N-1]$, and let $c:E\to [N-1]$ be an edge-coloring of $G$ as given in Lemma \ref{lemma-complete-coloring}. Next,  we define the retrieval scheme.  
	\begin{itemize}
	    \item[(a)] The user chooses a file index  $\theta \in [K]$ and a permutation $\sigma: [K]\to [K]$ uniformly at  random from the set of all permutations on the set $[K]$. 
	    \item[(b)] The user generates $N$ queries $Q_i$ for $i=1,\ldots, N$ and send them to the $N$ servers, as follows. To server $N$ it sends the permutation $\sigma$, whereas to server $1\leq  i< N$ it sends $\sigma\big(c(i,\theta)\big)$,  the  random color assigned to the edge $(i,\theta)$ by the random permutation $\sigma$ and the coloring $c$. Hence, the queries take the following form   
	    \begin{IEEEeqnarray*}{C}
	       Q_i:=Q_{i}(\theta,\sigma) = 
	       \begin{cases}
	       \sigma\big(c(i,\theta)\big) & 1\le i< N,\\ 
	       \sigma &  i=N.\\
	       \end{cases} 
	    \end{IEEEeqnarray*}
	    \item[(c)] Server $i$ for $i=1,\ldots,N$ replies with an  answer $A_{i}$ as follows. For $1\leq i<N$ the server returns the value of the $\sigma\big(c(i,\theta)\big)$-th bit of the file $W_i$ it holds, i.e., it returns $W_i\big(Q_i)$. On the other hand, server $N$ for each distinct $m,n\in [K]$  sends the sum of the $\sigma(c(m,n))$-th bits of the files $W_m$ and $W_n$. To conclude, the reply $A_i$ takes the form 
        \begin{IEEEeqnarray*}{C}
        A_i(Q_i)= 
            \begin{cases}
        	W_i\big(\sigma(c(i,\theta))\big) &  1\le i< N, \\
        	\Big\{W_m\big(\sigma(c((m,n)))\big) 
        	\oplus\>
        	    W_n\big(\sigma(c((m,n)))\big): \forall\,  m\ne n\in [K]\Big\} &  i=N.\\
        	\end{cases}
        \end{IEEEeqnarray*}
	\end{itemize} 
	Next, we show that the scheme has the claimed rate and satisfies the privacy and reliability requirements.
	
	\vspace{0.2cm}
	{\bf Rate:} Recall that the file size is $K$ bits. On the other hand, each of the first $N-1$ servers returns a single bit, and server $N$ returns  $\binom{K}{2}$ bits; therefore, the retrieval rate of the scheme is  
	\begin{equation*}
	    \frac{K}{\binom{K}{2}+(N-1)} =
	    \frac{2}{N}.
	\end{equation*}
	
	\vspace{0.2cm}
	{\bf Privacy:} Regardless of the value of $\theta$ and the coloring $c$, since  $\sigma$ is picked uniformly at random the $i$-th query $\sigma(c(i,\theta))$ for $i<N$ is uniformly distributed over $[K]$, and therefore it reveals no information on $\theta$. Similarly, the $N$-th query is completely independent of $\theta$ and therefore the scheme is private. 
	
	\vspace{0.2cm}
	{\bf Reliability:}  For any  $i\in [K]$, we will show that the user is able to retrieve the bit $W_\theta(\sigma(c(i,\theta)))$. Since, when $i$ ranges in $[K]$, $\sigma(c(i,\theta))$ ranges also in $[K]$ (here we use the property that $c$ is a proper coloring with $K$ colors), then the user is able to retrieve all the file bits, as needed. 
	
	For $i=\theta$ the user receives the bit $W_\theta(\sigma(c(\theta,\theta)))$ from server $\theta$. For $i\in [K],i\neq \theta$ the user   receives the bit $W_i(\sigma(c(i,\theta)))$ from server $i$, whereas from server $N$ it receives also the bit 
	$W_i(\sigma(c(i,\theta)))\oplus 
	        W_\theta(\sigma(c(i,\theta)))$. Therefore, by adding these two bits the user can recover the bit  $W_\theta(\sigma(c(i,\theta)))$, and the result follows.  
\end{IEEEproof}
	
    The above scheme uses a coloring of the complete graph and provides a rate that decays linearly with the number of servers in the system. Next,  we provide an improved scheme that uses the same idea of coloring; however, this time, the coloring will be of a hyper-graph, more specifically a Steiner system, defined below. The resulted scheme will be shown to be optimal up to a constant factor, i.e., it achieves the upper bound given in Theorem~\ref{thm-uppbd-star}. Hence, we have the following theorem. 
	
	\begin{theorem}\label{thm-lowbd-star-optimal}
	The rate of the star graph satisfies $\mathscr{C}(S_N) = \Theta(N^{-\frac{1}{2}})$.
	\end{theorem}
	The upper bound was proven in Theorem~\ref{thm-uppbd-star},	while the lower bound will follow from a PIR scheme with the claimed rate. As already mentioned, the improved scheme will rely on a coloring of Steiner system, whose definition is given next.
	
	\begin{definition}
	A Steiner system $\mathcal{S}(t,k,n)$ is a set system $(\mathcal{X},\mathcal{B})$, where $\mathcal{X}$ is an $n$-element set, say  $\mathcal{X}=[n]$, and 
	$\mathcal{B}\subseteq \binom{\mathcal{X}}{k}$ is a collection of $k$-element subsets (blocks) of
	$\mathcal{X}$, such that any $t$-element subset of
	$\mathcal{X}$ is contained in exactly one block of $\mathcal{B}$.
	\end{definition}
A Steiner system $\mathcal{S}(2,q,q^2)$ for a prime power $q$  can be constructed from $\mathcal{A}^2(\mathbb{F}_q)$ the \emph{classical affine geometry of dimension $2$}. Indeed, set   $\mathcal{X}=\mathbb{F}_q^2$, i.e., all the ordered pairs  pairs $(x, y)\in \mathbb{F}_q^2$ as elements; and blocks 
$\mathcal{B}\subseteq \binom{\mathcal{X}}{2}$
to be all subsets of the form $y = mx+b$ or $x = a$
(for fixed $a, m, b \in \mathbb{F}_q$), that is, blocks of the form 
\begin{equation*}
    \{(x, mx+b) : x \in \mathbb{F}_q\} \text{ for } m, b \in \mathbb{F}_q  \text{ or }   \{(a, y) : y \in \mathbb{F}_q \} \text{ for } a \in \mathbb{F}_q.
\end{equation*}

It is clear that there are total of $q^2$ elements in $\mathcal{X}$, the size of each block  $B\in \mathcal{B}$ is $|B|=q$ and that the number of blocks is $|\mathcal{B}|=q(q+1)$. In addition, for distinct elements   $x,y\in \mathcal{X}$ there exists a unique block, denoted by  $B_{x,y}\in \mathcal{B}$, that contains both $x$ and $y$ (this is simply the line that connects between $x$ and $y$).
Lastly, each element appears in exactly $q+1$ blocks.
    
	As we used an edge-coloring of a graph in the last construction, we will define a (proper block) coloring for a set system such that blocks that share an element have distinct colors. Formally, given a blocks set $\mathcal{B}$ of $\mathcal{X}$, a coloring of $\mathcal{B}$ with $T$ colors is a mapping $c:B\to [T]$ such that distinct colors are  assigned to distinct blocks that share an element, i.e., if blocks $B,B'\in \mathcal{B}$ are distinct blocks such that  $B \cap B' \neq \emptyset$, then $c(B)\neq c(B')$. Clearly, any coloring  of the Steiner system $\mathcal{S}(2,q,q^2)$ described above  requires at least $q+1$ colors, since  each  $x \in \mathcal{X}$ is contained in exactly $q+1$ blocks. The next lemma shows that the lower bound on the number of needed colors is tight, i.e., there is a coloring with only $q+1$ colors.

    \begin{lemma} \label{prop-block-coloring-lemma}
    The Steiner system $\mathcal{S}(2,q,q^2)$ admits a coloring with $q+1$ colors.
    \end{lemma}
	
	\begin{IEEEproof} 
	Each block of the Steiner system $\mathcal{S}(2,q,q^2)$ corresponds to a line in the affine plane   $\mathbb{A}^2(\mathbb{F}_q)$ with a slope   $m\in \mathbb{F}_q \cup \{\infty \}$, where we define the slope of a line parallel to the y-axis as $\infty$. 
	We let the color of a block be its slope $m$. One can easily verify that 
	every two distinct lines that share a point have distinct slopes; therefore,  this is a proper coloring of the blocks. 
	\end{IEEEproof}
    
    Next, we proceed to describe our scheme and prove its properties. 
	\begin{IEEEproof}[Proof of Theorem \ref{thm-lowbd-star-optimal}] 
    Recall that $S_N=(\mathcal{S},\mathcal{W})$ is a star graph with $N$ vertices where $S_i=\{W_{i}\}$ for any $1\le i\le N-1$, and $S_n=\mathcal{W}$. First, assume that the number of files $K=N-1=q^2$ for a prime power $q$, and suppose that each file $W=(W(1),\dots, W(q+1)) \in \mathbb{F}^{q+1}_2$ is a binary vector of length $q+1$.
	Let $(\mathcal{X},\mathcal{B})$ for $\mathcal{X}=[q^2]$ be the Steiner system  $\mathcal{S}(2,q,q^2)$ described above, and  
    let $c:B\to [q+1]$ be  the coloring as given in Lemma \ref{prop-block-coloring-lemma}.
    We define the retrieval scheme as follows.
    \begin{itemize}
	    \item[(a)] The user chooses a file index $\theta \in [q^2]$, a random index $\gamma\in [q+1]$  and a permutation $\sigma: [q+1] \to [q+1]$  uniformly at random from the set of all permutations on the set $[q+1]$.
	    
	    \item[(b)] The user generates $N$ queries $Q_i$ for $i=1,\ldots, N$ and send them to the $N$ servers, as follows. To server $N$ it sends the permutation $\sigma$, to server $\theta$ it sends $\gamma$, whereas to every other server $1\leq i < N, i\neq\theta$  it sends $ \sigma \left(  c \left( B_{i, \theta} \right) \right)$, the random color assigned to the block that contains $ i \text{ and } \theta $ by the permutation $\sigma$ and the coloring $c$. Hence, the queries take the following form 
        \begin{IEEEeqnarray*}{C}
       				Q_i : = Q_i(\theta, \gamma, \sigma) = 
            \begin{cases}
                      \sigma \left(  c \left( B_{i, \theta} \right) \right) & 1\leq i < N, i\neq \theta, \\
                      \gamma &  i=\theta, \\
                     \sigma &  i=N .\\
	        \end{cases}
        \end{IEEEeqnarray*}
	
	\item[(c)] Server $i$ for $i=1,\ldots,N$ replies with an  answer $A_{i}$ as follows. For $1\leq i<N$, the server returns the value of the $Q_i$-th bit of the file $W_i$ it holds. On the other hand, server $N$ for each block $B\in \mathcal{B}$ sends the sum of the $\sigma(c(B))$-th bits of all the files $W_j,j\in B$. Formally, the reply $A_i$ takes the form
    \begin{align*}
	        A_i (Q_i)= 
	        \begin{cases}
                      W_{i}(\sigma \left(  c \left( B_{i, \theta} \right) \right)) & 1\leq i < N, i\neq \theta, \\
                      W_{i}(\gamma) &  i=\theta, \\
                     \{ \bigoplus_{j\in B} W_j\left(\sigma(c(B)) \right)
                     :  \forall B\in \mathcal{B}  \}
                      &  i=N .\\
	        \end{cases}
	\end{align*}
	\end{itemize} 
	In what follows, we prove that the scheme has the claimed rate and it satisfies the privacy and reliability requirements.
    
    \vspace{0.2cm}
	{\bf Rate:} Recall that the file size is $q+1$ bits. Each of the first $N-1$ servers returns a single bit, while server $N$ returns $|\mathcal{B}|=q(q+1)$ bits, therefore the rate of the scheme is 
\begin{equation*}
	    \frac{q+1}{(N-1)+q(q+1)}\geq \frac{1}{2\sqrt{N-1}+1}
	    =
	    \Omega(N^{-\frac{1}{2}}),
\end{equation*}
	where we used  the fact that $q^2=N-1$.

	\vspace{0.2cm}
	{\bf Privacy:} Similarly to the privacy proof of Theorem \ref{thm-lowbd-star-simple}, for every $1\leq i \leq N-1$ the $i$-th query $Q_i$ is uniformly distributed in $[q+1]$ regardless of the value of $\theta$ and the coloring $c$, since  both $\sigma$ and $\gamma$ are chosen uniformly at random. For the same reason $Q_N$ is independent of $\theta$, hence the scheme is private.
	
	\vspace{0.2cm}
	{\bf Reliability:}  
	Let $B_1,\ldots, B_{q+1}$ be the $q+1$ blocks that contain  $\theta$. For any $i\in[q+1]$ 
	we will show that the user is able to retrieve the bit $W_\theta(\sigma(c(B_i)))$. 
	Note that  when $i$ ranges in $[q+1]$, $\sigma(c(B_i))$ ranges also in $[q+1]$ (here we use the property that $c$ is a proper block coloring with $q+1$ colors), then the user is able to retrieve all the file bits, as needed. 
	
	From each server $j\neq \theta$ such that $j \in B_i$, the user receives the bit $W_j({\sigma(c(B_i))})$, whereas from server $N$ it receives the bit 
	\begin{equation*}
	\bigoplus_{j\in B_i} W_j\left(\sigma(c(B_i)) \right).    
	\end{equation*}
	Then, by summing these $q$ bits, we have 
    \begin{equation*}
	    \bigoplus_{j\in B_i \setminus \{ \theta \}} W_j(\sigma(c(B_i))) \bigoplus_{j\in B_i} W_j(\sigma(c(B_i)))  =  W_{\theta}(\sigma(c(B_i))).
	\end{equation*}

Next, we address the case where the number of files $K$ is not a prime power. By the  Bertrand-Chebyshev theorem \cite{SR12}, let $q$ be a prime in the range   $\sqrt{K}\leq q \leq 2\sqrt{K}$, and define the scheme with $q^2$ files, where the files $W_i$ for $i=K+1,\ldots, q^2$ are dummy files, i.e., $W_i=0$. It is easy to verify that the rate of this scheme is also $\Omega(N^{-\frac{1}{2}})$, as needed.
\end{IEEEproof}

\begin{remark}
Although it is tempting to assume that one can not improve the scheme's rate by allowing different servers to transmit different amounts of information, this does not hold. In other words, it is possible to improve the scheme rate dramatically for certain graphs by allowing varying quantities of $H(A_i)$'s for different servers. Indeed, the above scheme for the star graph achieves the rate  of $\Omega(N^{-\frac{1}{2}})$ with non-constant answer size $H(A_i)$. On the other hand,  if one restricts the servers to transmit the same amount of information, i.e., 
$H(A_i)=H(A_j)$ for every $i,j\in[N]$, then 
by Theorem~\ref{theorem-exact-neighbors-bound} it follows that for any scheme of the star graph  $H(A_i) \ge L/2$, and therefore the rate is at most $2/N$, which is clearly much less than   $\Omega(N^{-\frac{1}{2}})$.

\end{remark}

\section{Retrieval schemes for the complete graph PIR system}\label{section-complete}
Understanding the complete graph's PIR capacity is of special importance since this graph contains the maximum number of edges (files) for a given number of servers. Furthermore, any graph $G$ on $N$ vertices is a subgraph of the complete graph $K_N$. Therefore any PIR scheme for $K_N$ can be converted into a scheme for $G$. In other words, it holds that $\mathscr{C}(K_N)\leq \mathscr{C}(G)$, which is a general lower bound for any graph $G$.

By  Remark~\ref{remark-complete-graph} the PIR capacity of $K_N$ is at most  $\frac{2}{N+1}$.
Whereas for the lower bound the following is known. In \cite{BU2018} it was shown that $\mathscr{C}(K_3)=\frac{1}{2}$ and $\mathscr{C}(K_4) \geq \frac{3}{10}$. For general $N$ we have $\mathscr{C}(K_N)\geq 1/N$.
In this section, we describe a general scheme that slightly improves the rate of $1/N$. More precisely,
we provide a scheme with rate $ \frac{2^{N-1}}{2^{N-1}-1}\cdot \frac{1}{N}$. Indeed, the improvement vanishes exponentially fast as the number of servers increases. However, the purpose of this construction is to provide an indication that the bound of $1/N$ is not optimal, and the upper bound of $\frac{2}{N+1}$ might be achievable with a more intricate scheme. 

We begin by describing the queries each server receives and the answer it provides in return; then, we explain how to generate the queries.
Recall that for distinct $i,j\in [N]$, $W_{i,j}$ is  the unique file stored on servers $i,j$, and  assume also that each file  is  a binary  vector of length  $L=2^{N-1}.$

\vspace{0.2cm}
\textbf{Queries:} Each  server $j$  receives  a  bijection 
\begin{equation}\label{stam}
\sigma_j:\{P:P\subseteq N(j)\}\rightarrow [L] 
\end{equation}
from the family of all subsets of its neighbors $N(j)$ to the set of integers $[L]$. 

\vspace{0.2cm}
\textbf{Answers:} Server $j$ returns $L-1$ bits, one for each nonempty subset of its neighbors, calculated as follows. For a nonempty subset $P\subseteq N(j)$ of its neighbors, the sum of all the $\sigma_j(P)$-th bits of files $W_{j,v},v\in P$ is computed, i.e., the bit 
\begin{equation}
\label{bit}
    b^j_{P}=\bigoplus _{v\in P} (W_{j,v})_{\sigma_j(P)}.
\end{equation}

\vspace{0.2cm}
\textbf{Generating the queries:} Assume that $W_{i,i'}$ is the requested file, and let $\Omega$ be the family of subsets of $[N]$ that contain exactly one of the servers $i$ or $i'$, i.e., 
\begin{equation*}
    \Omega=\{P\subseteq [N]: |P\cap \{i,i'\}|=1\}.
\end{equation*}
Clearly, $|\Omega|=L$,  and let $\pi:\Omega \to [L]$ be a bijection selected uniformly at random from the set of all possible bijections from $\Omega$ to $[L].$ 
Next, we describe how to construct for each server $j$ its bijection $\sigma_j$ as in \eqref{stam}. First, for  $j\neq i,i'$ let   $P\in \Omega$  be a set that contains $j$ and define 
\begin{equation}
\label{stam11}\sigma_j(P\backslash\{j\})=\pi(P).    
\end{equation}
Note that there are exactly $L/2$ such sets $P$. Extend $\sigma_j$ arbitrarily to a bijection as in \eqref{stam}, and note that the resulted bijection $\sigma_j$ is distributed uniformly among all the possible bijections. Therefore, server $j$ does not learn anything upon receiving its query. 

Next, for a server $a\in \{i,i'\}$ let $\overline{a}$ be such that $\{a,\overline{a}\}=\{i,i'\}$. 
Let   $P\in \Omega$  be a set that contains $a$ 
and  define 
\begin{equation}
\label{stam22}
\sigma_a(\{\overline{a}\}\cup P\backslash\{a\})=\pi(P),
\end{equation} and extend $\sigma_a$ arbitrarily to a bijection as in \eqref{stam}. Similarly, it is easy to verify that the resulted bijection $\sigma_a$ is distributed uniformly among all the possible bijections, and therefore, server $a$ does not learn anything upon receiving its query.
 
The following theorem shows that the above PIR scheme has the claimed properties. 

\begin{theorem}
\label{thm-completegraphs-imporved-lowerbd}
The above scheme is a PIR scheme with rate $\frac{2^{N-1}}{2^{N-1}-1}\frac{1}{N}.$  
\end{theorem}
\begin{IEEEproof}
\textbf{Reliability:}
We will show that the user can compute every bit of the file $W_{i,i'}$.  
For $l\in [L]$ let $P\in \Omega$ be the unique set such that $\pi(P)=l$ and let $a=P\cap \{i,i'\}$. Then, upon receiving the answers from the servers, the user can compute the following bit  
\begin{align}
    \bigoplus_{\substack{j\in P\\j\neq a}}b^j_{P\backslash \{j\}}\oplus b^a_{\{\overline{a}\}\cup P\backslash \{a\}}&=\bigoplus_{\substack{j\in P\\j\neq a}}\bigoplus_{v\in P\backslash \{j\}}(W_{j,v})_{\sigma_j(P\backslash \{j\})}\bigoplus_{v\in \{\overline{a}\}\cup P\backslash \{a\}}(W_{a,v})_{\sigma_a(\{\overline{a}\}\cup P\backslash \{a\})}\label{stam33}\\
    &=\bigoplus_{\substack{j\in P\\j\neq a}}\bigoplus_{v\in P\backslash \{j\}}(W_{j,v})_{l}\bigoplus_{v\in \{\overline{a}\}\cup P\backslash \{a\}}(W_{a,v})_{l}\label{stam44}
\end{align}
 where \eqref{stam33}  follows from \eqref{bit}; 
 and \eqref{stam44} follows from \eqref{stam11}, \eqref{stam22} and the fact that $\pi(P)=l$.
  Note that  the sum in \eqref{stam44} contains twice each  $l$-th bit of each   file  $W_{j,v}, j,v\in P$, and therefore is cancelled out, unless $\{j,v\}=\{a,\overline{a}\}$, where in this case it appears exactly once. Therefore, \eqref{stam44} is equal to $(W_{i,i'})_l$, as needed.

\vspace{0.2cm}
\textbf{Privacy:} As already mentioned,  each server receives a permutation distributed uniformly among all the possible permutations of the sets of subsets of its neighbors. Therefore, there is no information leakage. 

\vspace{0.2cm}
\textbf{Rate:} Each server return $L-1=2^{N-1}-1$ bits, hence the total amount of bits sent by all servers is $N\cdot (2^{N-1}-1)$, whereas the size of the file is  $L=2^{N-1}$. Therefore, the rate is $\frac{2^{N-1}}{2^{N-1}-1}\cdot \frac{1}{N}$, as needed.  
\end{IEEEproof}

The following example demonstrates how to construct the described scheme for $K_3$.

\begin{example} 
Let $\mathcal{S}=\{S_1,S_2,S_3\}$, $\mathcal{W}=\{W_{1,2},W_{1,3},W_{2,3}\}$ and suppose that each file $W\in \mathcal{W}$ is a vector in $\mathbb{F}_2^4$.
Without loss of generality assume that the required file is $\theta = W_{1,2}$ meaning that $\{i,i'\}=\{1,2\}$ and $\Omega = \{ \{1\}, \{2\}, \{1,3\}, \{2,3\} \}$. Assume that the randomly selected bijection $\pi: \Omega \rightarrow [L]$ is the bijection 
\begin{equation*}
  \pi(\{1\})=1,\pi(\{2\})=2, \pi(\{1,3\})=3,\pi(\{2,3\})=4.  
\end{equation*}
Next, we define the bijections $\sigma_j$'s.  Since $3\neq i,i'$,  then by \eqref{stam11}
\begin{equation*}
    \sigma_3(\{1\}) = \pi(\{1,3\})=3 , \sigma_3(\{2\})= \pi(\{2,3\})=4.
\end{equation*}
Next, for $\sigma_1$ note that $a=1, \bar{a} = 2$, hence by \eqref{stam22}
\begin{equation*}
  \sigma_1(\{2\})=\pi(\{1\})=1 \text{ and } \sigma_1(\{2,3\})=\pi(\{1,3\})=3.  
\end{equation*}
Similarly,
\begin{equation*}
  \sigma_2(\{1\})=2 \text{ and } \sigma_2(\{1,3\})=4.  
\end{equation*}
Finally, each of $\sigma_1,\sigma_2, \sigma_3$ is arbitrarily extended to a bijection 
 as in \eqref{stam}.  
Each server returns $3$ bits, but the user uses only two of them to retrieve the file $W_{1,2}$.
The needed bits from the three bits server $S_1$ returns are  
\begin{equation}
    b^1_{\{2\}}=W_{1,2}(1) \text{ and } b^1_{\{2,3\}}=W_{1,2}(3)\oplus W_{1,3}(3).
\end{equation}
Similarly, from server $S_2$, the user uses the bits   
\begin{equation*}
    b^2_{\{1\}}=W_{1,2}(2)\text{ and } b^2_{\{1,3\}}=W_{1,2}(4)\oplus W_{2,3}(4),
\end{equation*}
and from server $S_3$
\begin{equation*}
    b^3_{\{1\}}=W_{1,3}(3) \text{ and } b^3_{\{2\}}=W_{2,3}(4).
\end{equation*}
The user now can retrieve the file $W_{1,2}$ since  
\begin{IEEEeqnarray*}{C}
 W_{1,2}(1)=b^1_{\{2\}}, W_{1,2}(2)=b^2_{\{1\}},
 W_{1,2}(3)=b^1_{\{2,3\}} \oplus b^3_{\{1\}}, 
  \text{ and } 
 W_{1,2}(4)= b^2_{\{1,3\}} \oplus b^3_{\{2\}}.
\end{IEEEeqnarray*}
\end{example}

\section{Concluding remarks  and open questions}
\label{sec-conclusion}
In this paper, we studied the PIR capacity of graph-based replication systems with non-colluding servers. We proved several upper bounds that rely on the underlying graph structure, which turns out to be tight in certain cases. 
For lower bounds,  we used edge-coloring to establish PIR schemes for star graphs, which are optimal up to a constant factor. Also, we improved the known PIR schemes for complete graphs, which imply an improved lower bound for all graphs.

Despite these results, the PIR capacity of a graph is a parameter that is still far from being well-understood. Hence we conclude with several open questions. 
\vspace{0.2cm}

\setlist{nolistsep} 
\begin{enumerate}[noitemsep] 
    \item[1)]
    Improve the upper  (Corollary~\ref{remark-complete-graph}) and lower  (Section~\ref{section-complete}) bounds on the PIR capacity of complete graphs. 
    \vspace{0.2cm}
    \item[2)] Is it possible to extend the PIR  schemes for star graphs given in Section~\ref{section-star}  to other graph families, such as bipartite graphs, trees, regular graphs? 
    \vspace{0.2cm}
    \item[3)] The assumption that each file is stored on only two servers is too constraining; therefore, a natural question is whether the results given in this paper can be generalized to  
    hypergraphs. 
    
\end{enumerate}
\setlist{} 

\newpage
\appendix
\section{Proof of Lemma~\ref{lemma-symetric-scheme}} \label{proof-lemma-symetric-scheme}

For an arbitrary scheme $T$ and  a desired file with index $\theta$ (stored on servers, say  $k$ and $l$), we  denote by $Q_i^T(\theta)$ and $A_i^T$  the query sent to the  $i$-th server and its answer, respectively. Also,  we denote by $\mathcal{Q}^{T}$ the  set of queries generated by the user. 

Let $T$ be  a PIR scheme for a vertex transitive graph $G$. For an automorphism 
 $f\in Aut(G)$  of $G$,    let $f(\theta)$ be the index of the unique file  stored on servers $f(k),f(l)$.
 Given the scheme $T$ and the automorphism $f$, one can construct the scheme $T_f$ to retrieve the file with index $\theta$, as follows. The user sends to server $i$ the query $Q_i^{T_f}:=Q_{f(i)}^T(f(\theta))$ and the automorphism $f$, which in return replies with the answer $A_i^{T_f}:=A^{T}_{f(i)}$. 
 
 \vspace{0.1cm}
 {\bf Scheme analysis:} This scheme  can be thought of as applying the scheme $T$ on the graph $f(G)$ with a desired file with index $f(\theta).$ Note that the graph  $f(G)$ is simply a relabeling of every vertex $i$ by $f(i)$. Then,  by the reliability of $T$, the user can recover the file with index  $f(\theta)$, which is the unique file  stored  on vertices (servers) $f(k),f(l)$. Since we just relabeled the vertices, this is also the unique file stored on vertices $k,l$ in the graph $G$, i.e., the file with  index $\theta$, as needed. 
 
 The privacy requirement follows since we simply ran the original scheme $T$, which is assumed to be private, and clearly, the rate of the scheme $T_f$ is the same as of the scheme $T$.    
 
 \vspace{0.2cm} Next, we define below a scheme $T'$ that uniformly selects one of the schemes $T_f$ for  
 $f\in Aut(G)$ and uses it to retrieve the desired file. The exact details are as follows.  
\begin{enumerate}
\item[(a)] The user chooses a file $\theta\in[K]$, and an automorphism $f \in Aut(G)$ uniformly at random.
\item[(b)] The user generates the queries $\mathcal{Q}^{T'}:=\mathcal{Q}^{T_f}$, and sends them to the servers together with the automorphism $f$.  
\item[(c)] Each server $i$ replies with an answer $A^{T_f}_i$. 
\end{enumerate}

Clearly, $T'$ is a scheme having the same rate as $T$. 
Let $i,j$ be two servers; then by the transitivity of the automorphism group, there exists a $g\in \aut(G)$ with $g(j)=i$, then 
\begin{IEEEeqnarray}{rCl}
H(A_i^{T'} | \mathcal{Q}^{T'}) 
&=& \frac{1}{|Aut(G)|} \sum_{f\in Aut(G)} H(A^{T_{f}}_i | \mathcal{Q}^{T_f}) \label{eq-auto-f-uniform} \\
&=& \frac{1}{|Aut(G)|} \sum_{f\in Aut(G)} H(A^T_{f(i)} | \mathcal{Q}^{T}) \nonumber \\
&=& \frac{1}{|Aut(G)|} \sum_{f\in Aut(G)} H(A^T_{f(g(j))} | \mathcal{Q}^{T}) \nonumber\\ 
&=& \frac{1}{|Aut(G)|} \sum_{f\in Aut(G)} H(A^T_{f(j)} | \mathcal{Q}^T) \label{eq-auto-i-j} \\
&=& \frac{1}{|Aut(G)|} \sum_{f\in Aut(G)} H(A^{T_f}_{j} | \mathcal{Q}^{T_f})\nonumber\\
&=& H(A_j^{T'} | \mathcal{Q}^{T'}), \nonumber
\end{IEEEeqnarray}
where \eqref{eq-auto-f-uniform} follows since $f\in Aut(G)$ is chosen uniformly, and   \eqref{eq-auto-i-j} follows since if  $f$ ranges over all the automorphisms so does $fg$.  
This proves \eqref{lss-1}. 
Similarly, it can be shown that for any file $W_{i,j}$ 
 \begin{equation}
     H\big(A_i^{T'}|\mathcal{W}\setminus \{W_{i,j}\}) = H\big(A_j^{T'}|\mathcal{W}\setminus \{W_{i,j}\}). \label{eq-A-condition-i,j}
 \end{equation}
By Lemma~\ref{lemma-edge-bound}, we have 
 \begin{equation}
     H\big(A_i^{T'}|\mathcal{W}\setminus \{W_{i,j}\}) + H\big(A_j^{T'}|\mathcal{W}\setminus \{W_{i,j}\}) \geq L. 
     \label{ineq-sum-A-condition-i,j}
 \end{equation}
Combining~\eqref{eq-A-condition-i,j} and~\eqref{ineq-sum-A-condition-i,j} gives $H\big(A_i^{T'}|\mathcal{W}\setminus \{W_{i,j}\}) = H\big(A_j^{T'}|\mathcal{W}\setminus \{W_{i,j}\}) \geq \frac{L}{2}$. This completes the proof.

\end{document}